\documentclass[journal]{IEEEtran}
\usepackage{cite}
\usepackage{graphicx}

\ifCLASSINFOpdf
\else
\fi
\begin{document}
\title{Size-dependent Transport Study of In$_{0.53}$Ga$_{0.47}$As Gate-all-around Nanowire MOSFETs: Impact of Quantum Confinement and Volume Inversion}

\author{Jiangjiang~J.~Gu, Heng~Wu, Yiqun Liu, Adam T. Neal, Roy G. Gordon,
        and~Peide~D.~Ye% <-this % stops a space
\thanks{This work was supported in part by National Science Foundation and in part by Semiconductor Research Corporation (SRC) Focus Center Research Program (FCRP) Materials, Structures, and Devices (MSD) Focus Center. The authors would like to thank J. Shao, D. A. Antoniadis and M. S. Lundstrom for the technical assistance and valuable discussions.}
\thanks{J. J. Gu, H. Wu, A. T. Neal, and P. D. Ye are with the Department
of Electrical and Computer Engineering, Purdue University, West Lafayette,
IN, 47907 USA e-mail: (yep@purdue.edu).}% <-this % stops a space
\thanks{Y. Q. Liu is with the GlobalFoundries, Albany, NY, 12207 USA.}% <-this % stops a space
\thanks{R. G. Gordon is with the Department of Chemistry and Chemical Biology, Harvard University, Cambridge, MA, 02138 USA.}% <-this % stops a space
}

\maketitle

\begin{abstract}
%\boldmath
InGaAs gate-all-around nanowire MOSFETs with channel length down to 50nm have been experimentally demonstrated by top-down approach. The nanowire size-dependent transport properties have been systematically investigated. It is found that reducing nanowire dimension leads to higher on-current, transconductance and effective mobility due to stronger quantum confinement and the volume inversion effect. TCAD quantum mechanical simulation has been carried out to study the inversion charge distribution inside the nanowires. Volume inversion effect appears at a larger dimension for InGaAs nanowire MOSFET than its Si counterpart.
\end{abstract}


% \fi
\IEEEpeerreviewmaketitle

\section{Introduction}
\IEEEPARstart{I}{}nGaAs MOSFETs have recently been considered as one of the promising candidates for beyond 14nm logic applications~\cite{AlamoNature2011}. To meet the stringent demands of electrostatic control, non-planar 3D structures have been introduced to the fabrication of InGaAs MOSFETs, including InGaAs FinFETs~\cite{WuIEDMFinFET2009}, multi-gate InGaAs quantum-well FETs~\cite{RadosavljevicIEDM2010} and most recently, InGaAs gate-all-around (GAA) nanowire MOSFETs~\cite{GuIEDM2011}.

In particular, the InGaAs GAA nanowire MOSFETs have been shown to offer good scalability down to channel length ($L_{ch}$) of at least 50nm, thanks to the best electrostatic control of the GAA structure. High drive current ($I_{on}$) of 1.17mA/$\mu$m and peak transconductance ($g_{m}$) of 0.7mS/$\mu$m have also been achieved~\cite{GuIEDM2011} despite the non-optimized source/drain resistance ($R_{SD}$) and large equivalent oxide thickness (\emph{EOT}), showing great promise of the InGaAs GAA technology. Moreover, a detailed scaling metrics study has also revealed that reducing the nanowire size leads to improvements in subthreshold swing (\emph{SS}), drain induced barrier lowering (\emph{DIBL}), and threshold voltage ($V_{T}$) roll off, due to the tighter gate control~\cite{GuIEDM2011}. However, the impact of nanowire size on the transport properties of InGaAs GAA nanowire MOSFETs has not been studied and could lead to better understanding and design guidelines for next-generation InGaAs nanowire devices.

In this letter, we systematically investigate the impact of nanowire size on the on-state performance of InGaAs GAA nanowire MOSFETs. To our surprise, higher $I_{on}$ and intrinsic $g_{m}$ has been obtained on devices with smaller nanowire size. The low field mobility ($\mu_{0}$) is extracted using the Y-function method to further elucidate the transport performance of the nanowire devices~\cite{crosIEDM2006}, confirming the enhanced mobility for smaller nanowires. TCAD quantum mechanical simulation is employed to study the underlying physical mechanism~\cite{sentaurus}. It is shown that quantum confinement and volume inversion effects play an important role in the improved transport properties for the InGaAs GAA nanowire MOSFETs.

\section{Device Fabrication}
\begin{figure}[h]
  \centering
  \includegraphics[width=0.42\textwidth]{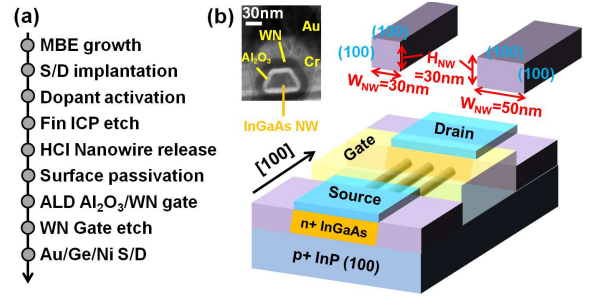}
  \caption{(a) Key fabrication process steps of InGaAs GAA nanowire MOSFETs by top-down approach. (b) Schematic diagram of an InGaAs GAA nanowire MOSFET and InGaAs nanowires with two different sizes under investigation (30nm$\times$30nm and 50nm$\times$30nm). Surface orientation (100) of the top surface and ideal side surface with vertical sidewall are illustrated. The nanowire patterning and current transport direction of [100] is depicted. A cross-sectional TEM image of an InGaAs GAA nanowire MOSFET with $W_{NW}$ of 30nm is also shown. The actual width and height of the nanowire is measured to be 26nm and 27nm with a $\sim$25$^o$ angle between actual and ideal sidewall. Note that nanowires with vertical sidewall and designed $W_{NW}$ is assumed in the normalization and simulation in this letter.}
  \label{Picture-1}
\end{figure}
Figure \ref{Picture-1} shows the fabrication process flow as well as the schematic diagram of the InGaAs GAA nanowire MOSFET. Fabrication started with a 30nm In$_{0.53}$Ga$_{0.47}$As channel layer with a p-type doping of $2\times10^{16} cm^{-3}$ epitaxially grown on a heavily p-doped InP (100) substrate by molecular beam epitaxy (MBE). After source/drain implantation (Si, $1\times10^{14} cm^{-2}, 20keV$), fin patterning was performed using BCl$_{3}$/Ar inductively coupled plasma (ICP) etching, followed by hydrogen chloride (HCl) based nanowire release process. The nanowires were aligned along [100] direction as required by the anisotropic HCl wet etching. After surface passivation with ammonia sulfide, 10nm Al$_{2}$O$_{3}$ and 20nm WN were grown by atomic layer deposition (ALD) at temperature of 300$^o$C and 385$^o$C respectively. Due to the excellent conformal coating ability of ALD, the gate stack forms surrounding all facets of the nanowires. Gate etch using CF$_{4}$ based ICP etching was then carried out to define the gate pattern. Finally, ohmic contacts were formed by electron beam evaporation of Au/Ge/Ni and liftoff process. Details of the fabrication process can be found in \cite{GuIEDM2011}.

The fabricated devices have nominal $L_{ch}$ varying from 120nm down to 50nm. Two different nanowire width ($W_{NW}$) (50nm and 30nm) were defined by lithography with a fixed nanowire height ($H_{NW}$) of 30nm defined by the MBE channel thickness. Since the nanowires were aligned along [100] direction, both the top and side surfaces are (100) surfaces assuming vertical sidewalls. Due to the non-optimized fin etching process, the actual sidewall leans 10 to 30 degrees towards (110) surface, confirmed by the SEM images~\cite{GuIEDM2011}.

\section{Results and discussion}
\begin{figure}[h]
  \centering
  \includegraphics[width=0.48\textwidth]{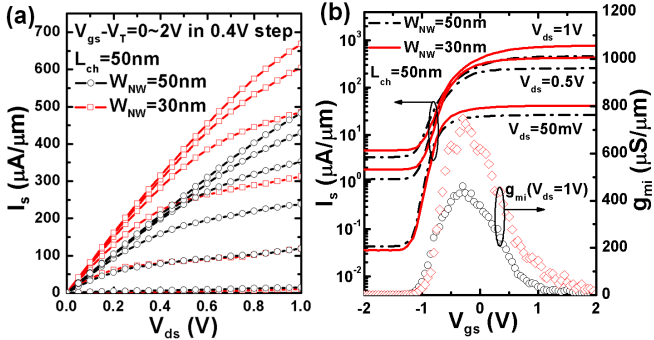}
  \caption{(a) Output characteristics and (b) transfer characteristics and intrinsic $g_{m}$ versus $V_{gs}$ of two typical InGaAs GAA nanowire MOSFETs with $L_{ch}=50nm$, $W_{NW}=30nm$ (red) and $50nm$ (black). $I_{s}$ is used due to relatively large junction leakage current.}
  \label{Picture-2}
\end{figure}
Figure \ref{Picture-2}(a) shows the output characteristics of two InGaAs GAA nanowire MOSFETs with $L_{ch}$ of 50nm. Devices with $W_{NW}$ of 30nm and 50nm exhibit a saturation current of 668$\mu$A/$\mu$m and 482$\mu$A/$\mu$m at $V_{gs}-V_{T}=2V$ and $V_{ds}=1V$, respectively. The current is normalized by the total \emph{perimeter} of the nanowires. The device with smaller nanowire size has a 38\% higher $I_{on}$. Similar enhancement in intrinsic $g_{m}$ is also observed on the smaller nanowire device as shown in Figure \ref{Picture-2}(b). The $V_{T}$ of the devices with $W_{NW}$ of 30nm and 50nm are $-0.85V$ and $-0.94V$ from linear extrapolation at $V_{ds}$=$50mV$, respectively. The negative $V_{T}$ is due to the low work function of WN ($\sim4.6eV$). Both devices show good pinch-off characteristics with a \emph{SS} of 150mV/dec at a $V_{ds}$ of 50mV. The upper limit of the interface trap density ($D_{it}$) at midgap is estimated to be 5.6$\times$10$^{12}$cm$^{-2}$$\cdot$$eV^{-1}$. The device with $W_{NW}$=30nm shows lower \emph{DIBL} ($\sim$180mV/V) compared to device with $W_{NW}$=50nm ($\sim$250mV/V), indicating improved control of short channel effects by shrinking the nanowire size. Considering the \emph{EOT} of $\sim$4.5nm and scaled $L_{ch}$ of 50nm, the \emph{SS} and \emph{DIBL} has been significantly improved compared to previous FinFET work~\cite{WuIEDMFinFET2009}, indicating the suitability of GAA structure for logic applications.

\begin{figure}[t]
  \centering
  \includegraphics[width=0.45\textwidth]{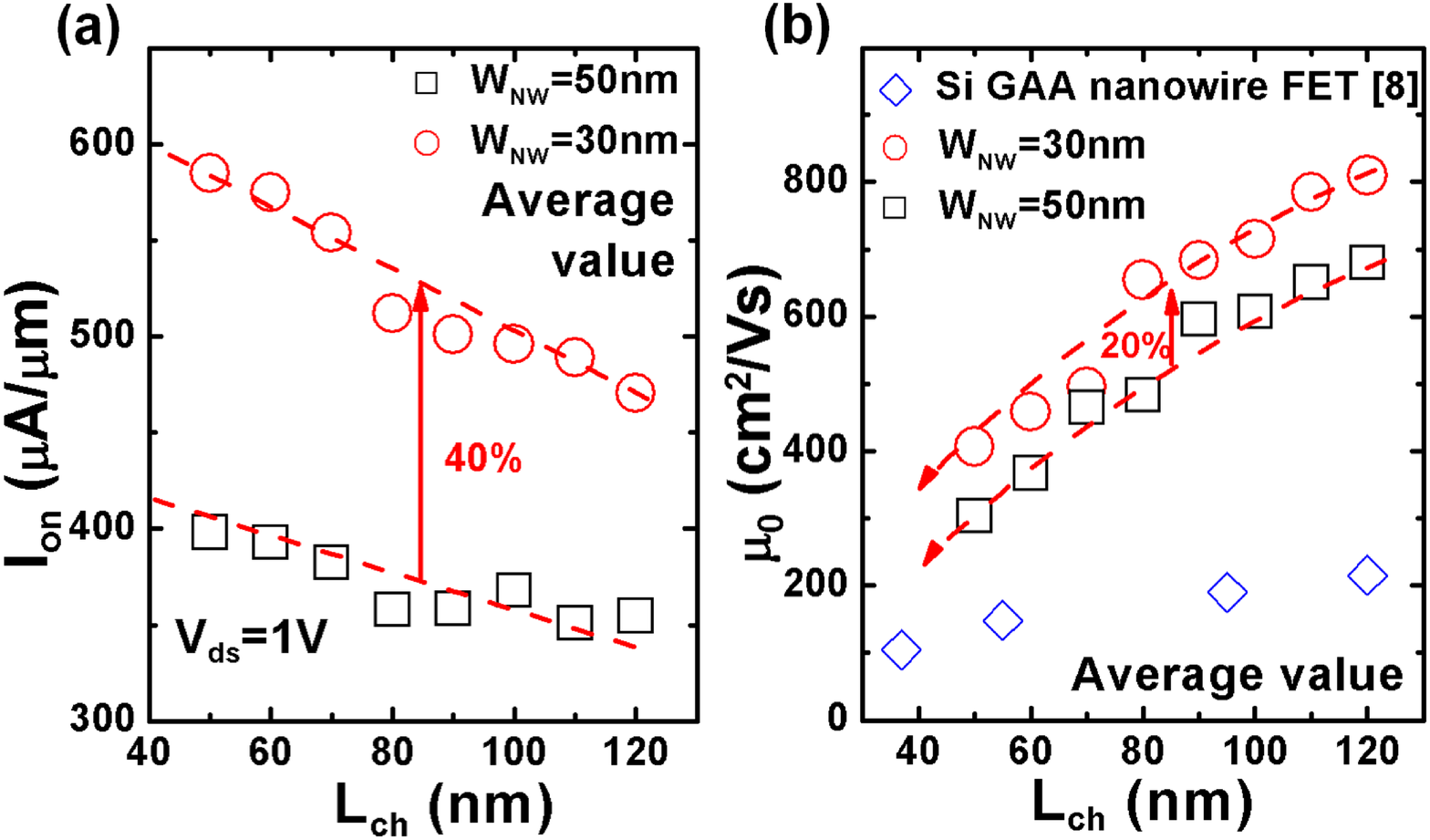}
  \caption{The average value of (a) $I_{on}$ measured at $V_{gs}-V_{T}=2V$ and $V_{ds}$=1V and (b) low field mobility $\mu_{0}$ of InGaAs GAA nanowire MOSFETs extracted using the Y-function method with $L_{ch}$ varying from 50nm to 120nm and $W_{NW}$ of 30nm (red circle) and 50nm (black square), compared to Si GAA nanowire NMOSFETs data~\cite{RWangEDL2008} (diamond). Over 20 devices were measured at each data point to obtain the average value.}
  \label{Picture-3}
\end{figure}

Figure \ref{Picture-3}(a) shows the average $I_{on}$ measured at $V_{gs}-V_{T}=2V$ and $V_{ds}=1V$ as a function of $L_{ch}$. A gradual increase of $I_{on}$ is observed when scaling down the channel length for both nanowire sizes. An average of 40\% increase in $I_{on}$ has been obtained on devices with $W_{NW}$ of 30nm over the entire $L_{ch}$ range. Devices with different $W_{NW}$ show similar $R_{SD}$, ranging from 950 to 1150$\Omega\cdot\mu$m. The intrinsic $g_{m}$ of devices with smaller nanowire size is found to be 34\% higher than those with larger nanowire size (not shown). To further characterize the transport in the nanowire devices, effective mobility was extracted using the Y-function method, which agrees reasonably well with the split-CV method and allows for the suppression of the series resistance effect~\cite{crosIEDM2006}. Figure \ref{Picture-3}(b) shows the average $\mu_{0}$ versus $L_{ch}$, demonstrating over 20\% mobility enhancement for devices with smaller $W_{NW}$. The apparent mobility reduction at shorter $L_{ch}$ can be explained by Shur's model using a Mathiessen-like relation considering the ballistic mobility~\cite{ShurEDL2002}. It is also shown in Figure \ref{Picture-3}(b) that the extracted $\mu_{0}$ of the InGaAs GAA nanowire MOSFETs are \emph{2$\sim$4 times} higher than those from state-of-the-art Si nanowire devices~\cite{RWangEDL2008}, owing to the better transport properties of the III-V channel.

The increase in $I_{on}$, $g_{m}$ and $\mu_{0}$ has confirmed that improved transport has been obtained in smaller InGaAs nanowires. Normally for top-down nanowires, it is expected that reducing the nanowire size will degrade transport due to the relative increase in surface roughness scattering given the larger surface to volume ratio of the ultra-small nanowires. However, it has been reported on Si nanowire MOSFETs that the improved transport from high-mobility sidewall~\cite{JZChenEDL2009}, oxidation induced strain inside the nanowire~\cite{MoselundTEM2010}, and the volume inversion effect in nanowires with small cross sectional area~\cite{SukIEDM2007} would result in enhanced transport properties with $W_{NW}$ shrinkage. The InGaAs (111)A surface has been demonstrated to offer higher mobility than other crystal orientations due to the trap redistribution~\cite{IshiiAPEX2009}. However, (111)A surface can not be the sidewall facet of InGaAs nanowires in this study, since the nanowires are aligned along [100] direction. Moreover, the thermal budget of the fabrication process after nanowire release in this letter is as low as 385$^{o}$C, which is much lower than the thermal oxidation temperature (usually over 1000$^{o}$C) of the Si nanowire MOSFET~\cite{MoselundTEM2010}. Therefore, strain-induced mobility enhancement can not play a significant role in the InGaAs nanowires under investigation. On the other hand, due to the much smaller effective mass and density of states of InGaAs, the inversion layer thickness can be 3.5 times larger than that of Si. As a result, inversion carriers can be pushed further away from the interfaces due to a stronger quantum confinement leading to the volume inversion effect in InGaAs nanowires at larger dimensions than Si.

\begin{figure}[h]
  \centering
  \includegraphics[width=0.425\textwidth]{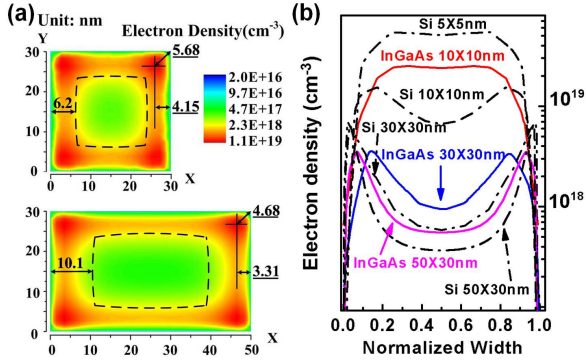}
  \caption{(a) Cross sectional distribution of electron density in the In$_{0.53}$Ga$_{0.47}$As nanowire with $W_{NW}$ of 30nm and 50nm at $V_{gs}-V_{T}=1.2V$. Dashed line shows the inversion layer centroid. (b) Normalized electron distribution at the middle of the nanowire ($y=0.5H_{NW}$) for square-shape Si and In$_{0.53}$Ga$_{0.47}$As nanowires. Note that vertical sidewalls are assumed in the simulation. While the sidewalls in the experiments are not vertical, resulting in reduced gate control, the general scaling trend comparing nanowires with different sizes remains unchanged.}
  \label{Picture-4}
\end{figure}

To further clarify the underlying mechanism, TCAD simulation using Sentaurus Device~\cite{sentaurus} was carried out. The electron distribution inside the nanowire in the strong inversion regime is obtained using a coupled Poisson and quantum potential solver based on the density gradient model~\cite{sentaurus, AnconaPRB1987}, considering only $\Gamma$ valley for InGaAs. It is found that both InGaAs GAA nanowire MOSFETs with $W_{NW}$ of 30nm and 50nm operate in the volume inversion regime, where the inversion charge density inside the entire nanowire is higher than the background p-type doping. The $W_{NW}$=30nm device show stronger confinement, resulting in the inversion layer being pushed 1$\sim$2nm further away from the surface and a higher inversion charge density at the center of nanowire compared to the $W_{NW}$=50nm case, as shown in Figure \ref{Picture-4}(a). This would lead to suppressed surface roughness scattering for the smaller nanowire. Furthermore, the volume inversion also results in the inversion layer centroid of the smaller nanowire being closer to the surface and therefore an increase in electrostatic capacitance with decreasing $W_{NW}$. It is also found that the two inversion layers inside InGaAs nanowire would merge into one peak at a dimension of $\sim$10nm, twice as large as that in the Si case ($\sim$5nm) as shown in Figure \ref{Picture-4}(b). The inversion layer distributes further inside the InGaAs nanowire with a higher density at the center compared to the Si case with the same nanowire size. Further experimental efforts reducing InGaAs nanowire size are required to illuminate on the ultimate scaling limit of InGaAs GAA nanowire MOSFETs, which may require development of new nanowire thinning techniques. The volume inversion at a larger dimension and a stronger quantum confinement in the InGaAs GAA nanowire MOSFETs may relax the fabrication complexity and interface quality requirements for InGaAs nanowire devices.

\section{Conclusion}
In this letter, we have fabricated and characterized InGaAs GAA nanowire MOSFETs with different nanowire size. Enhanced transport properties have been confirmed on InGaAs nanowires with a smaller dimension, due to a stronger quantum confinement and volume inversion effect. It is shown that distribution of inversion carriers moves further away from the surface and volume inversion occurs at a larger dimension on InGaAs nanowire than its Si counterpart, making InGaAs GAA MOSFETs favorable for future logic applications.

% or


%\appendices

%\section{}

%\section*{Acknowledgment}

%The authors would like to thank R. Wang, M. Luisier, M. S. Lundstrom, D. A. Antoniadis, J. A. del alamo, C. Zhang, and X. L. Li for valuable discussions and technical assistances. This work is supported by the SRC FCRP MSD and NSF.

\ifCLASSOPTIONcaptionsoff
  \newpage
\fi




\bibliographystyle{IEEEtran}
\end{document}